\documentclass[12pt,a4paper]{article}
\usepackage[hscale=0.8,vscale=0.85]{geometry}
\usepackage{graphicx}

\begin{document}
 
\title{Entropy and Quantum Gravity}

\author{Bernard S. Kay\\
\vspace{0.3 true cm}
{\small Department of Mathematics, University of York, York YO10 5DD, UK}\\
{\small \tt bernard.kay@york.ac.uk}}

\date{}

\maketitle

\begin{abstract}
We give a review, in the style of an essay, of the author's 1998 matter-gravity entanglement hypothesis which, unlike the standard approach to entropy based on coarse-graining, offers a definition for the entropy of a closed system as a real and objective quantity.   We explain how this approach offers an explanation for the Second Law of Thermodynamics in general and a non-paradoxical understanding of information loss during black hole formation and evaporation in particular.  It also involves a radically different from usual description of black hole equilibrium states in which the total state of a black hole in a box together with its atmosphere is a pure state -- entangled in just such a way that the reduced state of the black hole and of its atmosphere are each separately approximately thermal.  We also briefly recall some recent work of the author which involves a reworking of the string-theory understanding of black hole entropy consistent with this alternative description of black hole equilibrium states and point out that this is free from some unsatisfactory features of the usual string theory understanding.   We also recall the author's recent arguments based on this alternative description which suggest that the AdS/CFT correspondence is a bijection between the boundary CFT and just the matter degrees of freedom of the bulk theory.
\end{abstract}

\bigskip

What follows is a review, in the style of an essay, of the author's matter-gravity entanglement hypothesis -- recalling the basic issues it addresses and its basic statement.  It includes a discussion of some more recent work of the author on how existing results on string theory and the AdS/CFT correspondence related to black hole equilibrium states can be reworked and reconciled with our hypothesis.   We argue that the resulting reworking leads to a clearer understanding than hitherto available of black hole entropy in terms of string theory and a clearer resolution to the information loss puzzle and also to some other puzzles including the Arnsdorf-Smolin puzzle \cite{ArnsdorfSmolin} related to AdS/CFT.   Our purpose is to collect together, in a single short and easily readable article, the main evidence obtained so far for the validity of our hypothesis.   The emphasis is on the main ideas.  The referenced papers by the author should be consulted for full details.

The Second Law of Thermodynamics originated with Carnot (1828) as a statement about which changes of state are possible for machines such as heat engines and refrigerators.  One way to state it is: 

\medskip
\noindent
\emph{The entropy of a closed system always increases with time.}

\medskip

Entropy (the term was coined by Clausius in 1856) was originally defined in terms of the macroscopic phenomenological quantities, `heat' and `temperature'.   But to go beyond systems which depart from thermal equilibrium only slightly or at a slow rate, one needs a more fundamental definition:  

In a classical setting, Boltzmann's 1877 proposal was (in the terminology of Planck) that entropy ($S$) equals Boltzmann's constant ($k$) times the logarithm of the number ($W$) of microstates belonging to a given macrostate:

\[
S=k \log W.
\]

   This equation was subsequently adapted to a quantum mechanical setting by von Neumann with the formula 

\[
S = - k\, {\rm tr}(\rho\log\rho).
\]

where $\rho$ is the system's coarse-grained density operator.  And yet, as von Neumann famously remarked in a conversation with Shannon in 1948, ``nobody knows what entropy really is''.      

Amongst the reasons entropy may seem a mysterious and elusive concept are, firstly, that there seems to be a danger of a contradiction between the time-irreversible Second Law and the time-reversal invariant (or at least PCT invariant) microscopic laws of physics. Secondly, the process of coarse-graining by which we group together microstates into equivalence classes of macrostates to define the Boltzmann entropy is necessarily partly arbitrary, based as it is on a subjective judgement about which pairs of states are indistinguishable. 

One might argue that none of this matters and entropy is not a fundamental quantity; the only truly fundamental and natural value for the $W$ in Boltzmann's 
formula is 1 (all distinct microstates are ultimately distinguishable) and the only natural value for the entropy, $S$, of any state of any closed system is therefore zero.  Likewise, a full description of a quantum closed system would be with a \emph{pure} density operator, for which the von Neumann entropy is again zero.  

And yet, there seem to be reasons \cite{Davies, Penrose} to believe that the universe really does have a non-zero entropy -- and that this is quite independent from any subjective judgments that we may make about what we can and cannot distinguish\footnote{It is interesting to note, in relation to the entropy of the universe, that, in \cite{Lieb}, while commenting on the failure of monotonicity of entropy for quantum systems, Lieb remarks that it ``presents $\dots$ a problem for physics $\dots$ that $\dots$ the entropy of our planet could increase without limit while the entropy of the universe remains zero.''   He then goes on to indicate that this dilemma can be resolved in view of general theorems which ensure that, for large enough quantum systems, entropy is approximately additive.  We wish to remark here, in relation to this, that (a)  it is not clear whether such general theorems on the approximate additivity of entropy are applicable in quantum cosmology (b) were approximate additivity to hold and on the assumption (which Lieb and others make) that the entropy of the universe is the von Neumann entropy of its total density operator, the resolution proposed by Lieb would entail a total state of the universe which is very far from pure.   Our proposal to identify the total entropy of the universe, instead, with the matter-gravity entanglement entropy of the total density operator offers the prospect of a quantum cosmology with a total pure state and yet a large and increasing entropy for the universe.}.   Indeed its value has been estimated (see e.g.\ \cite{Framptonetal}).   Furthermore, thanks to Hawking \cite{HawkingEvap}, we know that a black hole has an entropy equal to a quarter of the area of its event horizon -- and there certainly seems to be nothing subjective about a quarter of an area!  Moreover, presumably the entropy of the universe really is increasing and  the entropy of a model closed system consisting of a star in empty space which collapses to a black hole and subsequently Hawking-evaporates will (when we include the contribution to the entropy from the radiated particles) increase monotonically with time.

In 1998 I made a proposal \cite{KayEntr, KayDeco, KayAbyaneheeee} as to what the connection between the microscopic laws of physics and the laws of thermodynamics might be according to which the entropy of a closed system is a real and objective quantity. With this proposal,  the question of whether entropy increases monotonically with time becomes, with suitable assumptions about the microscopic laws of physics and suitable assumptions about initial conditions, a well-defined and meaningful mathematical question.  As for what those microscopic laws of physics are, we don't need to say in detail to see how the proposal might work.  All we need to assume is that there is an approximate \emph{quantum gravity} theory valid for energies well below the Planck energy and that this can be formulated along the lines of a standard quantum mechanical theory with a total Hilbert space, $\cal H$, which arises as the tensor product of a matter Hilbert space, ${\cal H}_{\mathrm{matter}}$, and a gravity Hilbert space, ${\cal H}_{\mathrm{gravity}}$, together with a unitary time-evolution for an ever pure total density operator\footnote{These very basic assumptions are all that is needed to state our matter-gravity entanglement hypothesis and hence, for clarity, we have deliberately refrained at this point in the main text from mentioning any particular proposals for such approximate quantum gravity theories.    Also we have refrained from attempting to indicate how such an effective description could emerge from a fundamental theory.  However we remark: (a) We have explored one simple low-energy approximation -- a version of `Newtonian quantum gravity' in \cite{KayDeco, KayAbyaneheeee} -- see also \cite{KayAbyanehRobust}.   It would be interesting to attempt to investigate what other versions of low energy quantum gravity such as those reviewed in \cite{Donoghue} might imply for matter-gravity entanglement; (b) in the context of the semi-qualitative string theory understanding of black hole equilibrium states due to Susskind \cite{Susskind} and Horowitz and Polchinski \cite{HorowitzPolchinski, HorowitzChandra}, we arrived, in \cite{KayModern, KayStringy}, at a tentative understanding of how  ${\cal H}_{\mathrm{matter}}$ and ${\cal H}_{\mathrm{gravity}}$ arise as emergent features from more fundamental string-theory degrees of freedom.  (See especially the paragraph containing Equation (7) and the subsequent paragraph in \cite{KayStringy}.) We recall some of the main ideas of this work below.  We remark here that one lesson from our work in \cite{KayModern, KayStringy} is that, to understand quantum black holes, while the states of quantum gravity which are relevant may be `low energy' states, the coupling between matter and gravity cannot be regarded as weak.   Moreoever, if we assume the theory can however be equated to a weakly coupled string theory in the weak string-coupling limit, then the densities of states, $\sigma(\epsilon)$,  of what `matter' and `gravity' go over to in that limit will not  be of the ordinary power law sort, $\sigma(\epsilon)) \sim C \epsilon^N$ for some large number $N$ (here we use $\epsilon$ to denote energy)  but rather grow exponentially with energy (with a prefactor which is an inverse power of energy).}
 
In such a theory, the von Neumann entropy of the total state will of course be zero at all times.  But, and this is the crucial new feature of the proposal, we don't identify the physical entropy of the total state with its von Neumann entropy.   Rather \cite{KayEntr, KayDeco, KayAbyaneheeee} we identify it with the total state's \emph{matter-gravity entanglement entropy} -- i.e.\ with the von Neumann entropy of the reduced density operator for the matter -- obtained by taking the partial trace of the total pure density operator over the gravity Hilbert space (which, since the total state is pure, happens, by a well-known easy theorem, to equal the von Neumann entropy of the reduced density operator for gravity -- obtained by taking the partial trace of the total pure density operator over the matter Hilbert space).  There is no reason why this quantity should remain zero for all time and indeed, with the further assumption that the initial state of the closed system has a low degree of matter-gravity entanglement, it is plausible that it will increase monotonically for all time. 

Thus we have a plausible explanation for the Second Law for a general closed system.   Applied to our collapsing star closed system, and bearing in mind that information may be defined as negative entropy, this specializes to a (non-paradoxical) explanation of how information is lost in black-hole collapse.  So we see that, on our view, the `information-loss puzzle' \cite{HawkingInfoLoss} is  just a special instance of the more general puzzle of how, for any closed system, its entropy increase can be reconciled with a unitary time evolution.  Once one ceases to identify the physical entropy of the closed system with the von Neumann entropy (a unitary invariant) of its total state and identifies it instead with the total state's matter-gravity entanglement entropy, both the general puzzle and its special case, which relates to black holes, go away.

Our proposal resembles the environment induced decoherence paradigm \cite{Zurek} but with a crucial difference:  In the environment paradigm, one separates one's total closed system into a `subsystem of interest' and an `environment' and regards the subsystem-environment entanglement entropy as the physical entropy of the (\emph{open}) subsystem.   But in our proposal, the matter-gravity entanglement entropy is identified with the entropy of the total closed system -- and will, in general, be non-zero even though the state of the total closed system is, at all times, a pure state!  

Our proposal can easily be extended to include both closed and open systems by realizing the matter Hilbert space as a tensor product of a matter-system and a matter-environment Hilbert space:  In some given closed matter-gravity system in some given total pure state, we then define the entropy of some given open subsystem of the matter to be the von Neumann entropy of its reduced density operator -- obtained by taking the partial trace of the total density operator over the appropriate matter-environment Hilbert space as well as over the gravity Hilbert space (i.e.\  by taking the partial trace over the tensor product of the latter two Hilbert spaces).   See Endnote (xii) of \cite{KayAbyaneheeee} for details.  Now as one considers increasing the size of what we consider to be the matter system, and concomitantly reducing the size of what we consider to be the matter environment -- schematically indicated by sliding the vertical dotted line to the right in Figure 1a, one expects the entropy of the matter system to tend, in the limit as one slides it fully to the right, to the non-zero value for the entropy of the total closed system -- as in the schematic graph in Figure 2a.   This is to be contrasted with what would happen on the standard environment paradigm (on the assumption of a total pure state) schematically illustrated by sliding the dividing line between system and environment to the right in Figure 1b.  The entropy may increase at first, but eventually it must decrease towards the value zero for the total closed system as in the schematic graph in Figures 2b.  (See also Footnote 1.)

\begin{figure}
   \centering
    \includegraphics*[scale=1.00, trim= 60 500 0 100, clip]{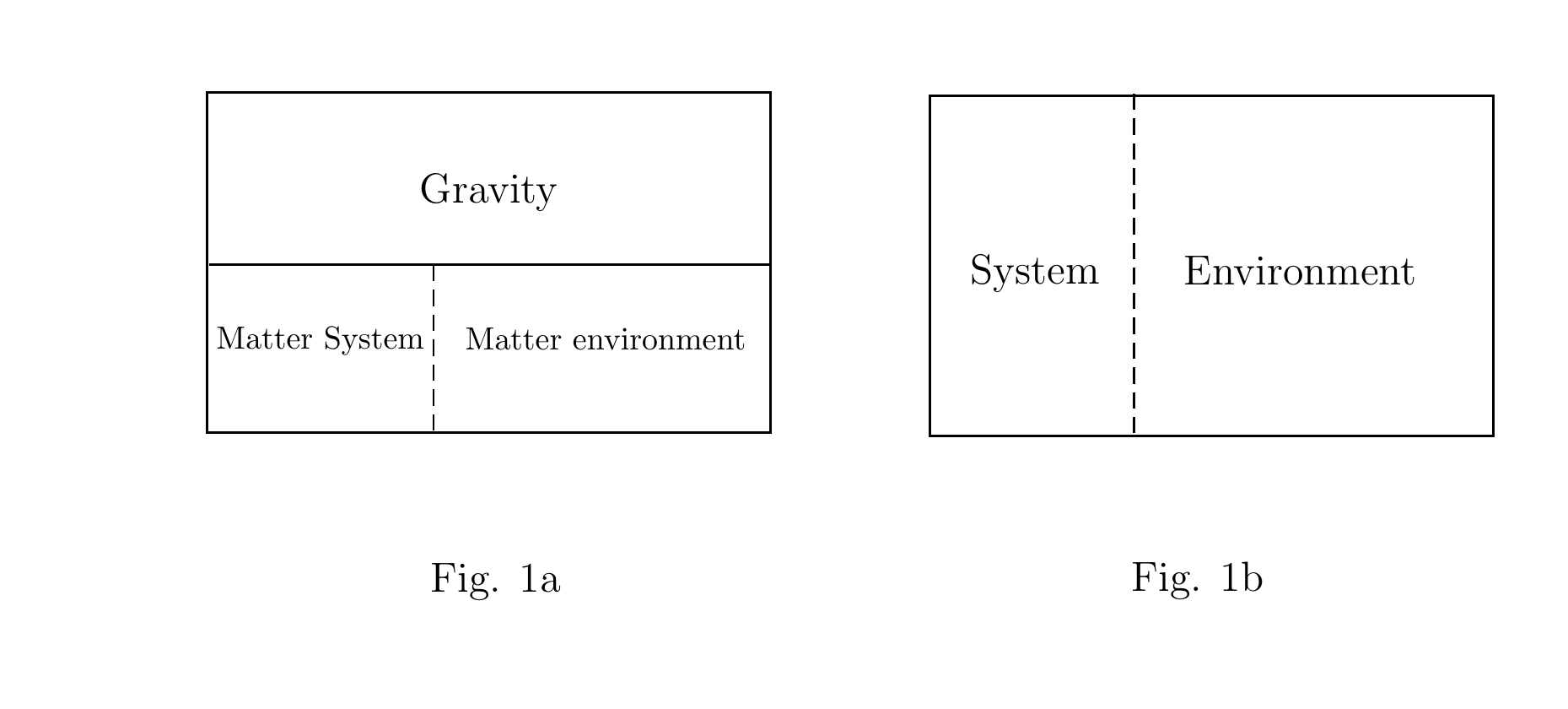}
   \caption{Schematic diagrams contrasting our approach to open systems (Fig. 1a)
with that on the traditional `environment-induced decoherence' paradigm (Fig. 1b).
 \label{fig1} }
 \end{figure}

\begin{figure}
     \centering
    \includegraphics*[scale=0.90, trim=120 450 0 100, clip]{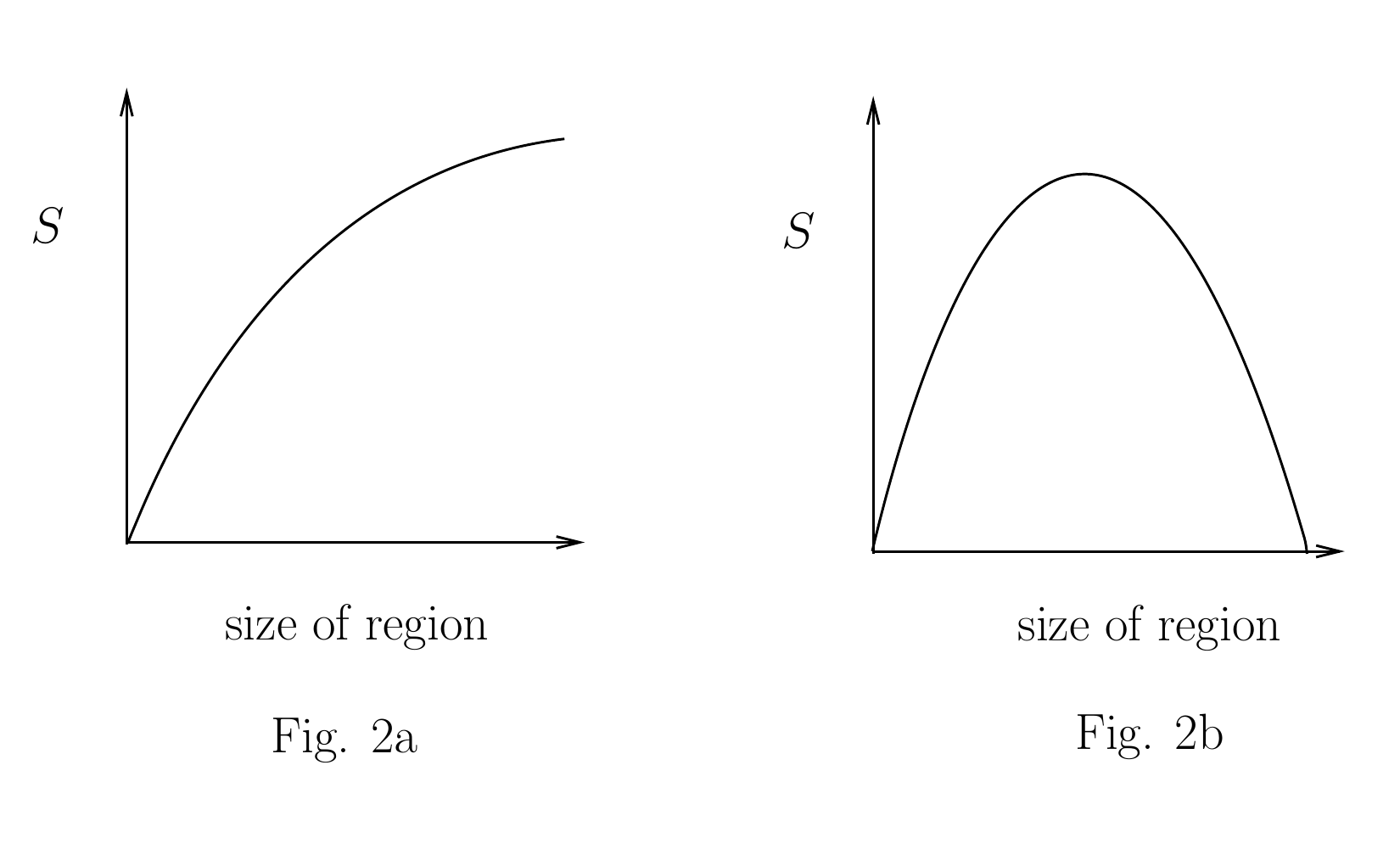}
   \caption{Schematic behaviour of entropy against `size of open system' for our
approach to open systems (Fig. 2a) contrasted with the corresponding behaviour
on traditional ideas (Fig. 2b). (We note that Figures 1 and 2 first appeared as `Figure 4' in Reference \cite{KayAbyaneheeee}.) \label{fig2}}
 \end{figure}

One expects that, in accounting for the entropy of ordinary macroscopic open subsystems of matter with typical laboratory sizes and energies (such as gases in boxes etc.)\ or indeed typical terrestrial sizes and energies, it won't make any significant difference whether one neglects gravity and regards the entropy of some such system as arising from tracing over the matter environment or whether one includes gravity in the theory and traces over the matter environment as well as the gravitational field as we propose.   This is an important check that our matter-gravity entanglement proposal is reasonable.   Had we instead  based our hypothesis on the factorization of the total Hilbert space as a tensor product, say, between a Hilbert space for matter except for the electromagnetic field and  a Hilbert space for gravity together with the electromagnetic field, then (since the electromagnetic field is such an important component of ordinary matter) one would no longer be able to view the theory of the origin of the entropy of ordinary macroscopic systems provided by the usual environment paradigm as a limiting case of our new theory. 
However, our main argument for basing our hypothesis on the factorization of ${\cal H}$ as ${\cal H}_{\mathrm{matter}}\otimes {\cal H}_{\mathrm{gravity}}$ is based on the fact that it offers a resolution to the thermal atmosphere puzzle, as we discuss below.

Our above proposed explanation of the Second Law relies on our closed system having an initially low entropy (i.e.\ low degree of matter-gravity entanglement).   If a supposedly closed system (e.g.\ our collapsing star model system) is really an approximately closed part of a bigger universe, this low entropy will (cf. \cite{Davies}) presumably be traceable to a low initial entropy of the universe as a whole.  We don't explain why that might be low, but one might hope that a more ambitious theory might do so.    

In the traditional approach, one can similarly \cite{Davies, Penrose} reconcile irreversibility with reversible microscopic laws by assuming a low entropy initial state for the universe.   But the explanation will inherit an unsatisfactory subjective aspect due to the subjective nature of entropy as traditionally understood.  To make this clear, consider, e.g.\ the classic thought experiment where one removes a partition separating two equal halves of a, say, rectangular box containing a single non-relativistic particle, initially confined, say, to the left half of the box (say with a pure-state wave function satisfying vanishing boundary conditions at the walls of the box and at the partition). The initial state is deemed to have a lower entropy -- by a factor of $k\log 2$ -- than the state after the removal of the partition but only because we declare ourselves able to distinguish between a state where the particle is definitely confined to one half of the box and a state where all we know about it is that it is located somewhere in the entire box with (one expects) roughly equal probabilities of being in the left and right halves.  

In the traditional (von Neumann \cite{Neumann}) approach to quantum mechanics, one can understand this traditional explanation of the increase of entropy in this box model as due to the performance of a quantum mechanical measurement to answer the question whether the particle is in the left or right half of the box.  Before the partition is removed, the measurement outcome will be that the particle is definitely in the left half of the box and the density operator will remain the (pure) projector onto the initial wave function -- with von Neumann entropy zero.   At most times after the partition is removed, the density operator after the measurement will be a mixture, with roughly equal probabilities, of a projector onto a wave function localized in the left half of the box and a projector onto a wave function localized in the right half of the box; and this density operator will, of course, therefore have von Neumann entropy (approximately) equal to $k\log 2$.  The unsatisfactory subjective element in the understanding of entropy (and of entropy increase) in this quantum mechanical version of the traditional approach is the fact that it needs to refer to the notion of `measurement'.  In fact we see that the unsatisfactory \cite{Bell} subjective aspect of the traditional approach to quantum mechanics and the unsatisfactory subjective aspect of the traditional understanding of entropy that we are focussing on in this essay are closely interrelated.

On the usual environment-induced decoherence paradigm, one overcomes this unsatisfactory subjectiveness and arrives at an objective notion of entropy, but only for open systems.   If the entire box in our above box model is deemed to be an open system, coupled (even if perhaps only very weakly) to an environment in such a way that the system together with the environment is in an overall pure state, then one expects that, when the partition is removed, the entropy of the box -- now understood as its entanglement entropy with its environment -- will rapidly increase by the same factor of $k \log 2$ predicted on the above traditional approach.  However,  if such a box were truly a closed system, without any environment,  then, in the spirit of the environment approach where one does not admit the occurence of measurements, one would say that its entropy would not increase on removing the partition, the state remaining at all times pure.  

The extension of our matter-gravity entanglement proposal to open systems, mentioned above, would lead to essentially the same conclusion for the removal of the partition in our box model when it is coupled to a matter environment.  However, as we have already explained above, our matter-gravity entanglement hypothesis also predicts, as an objective fact, the increase of entropy for closed systems.  In particular, for our box model, one expects it would predict an increase in entropy,  even if our box were a truly closed system, with no matter environment -- albeit this increase might be very small.  This is because, on our matter-gravity entanglement hypothesis, we would include the gravitational field in the description of our total system and we would not neglect the interaction of the particle with gravity (even though, for many other questions, the effects of that coupling might be quite negligible) and we would equate the total system's entropy with its matter-gravity entanglement entropy.   

Our conclusions from the above four paragraphs are:  The traditional approach to entropy and to entropy-increase applies to closed systems, but (at least when it is applied to closed systems) it has an unsatisfactory subjective element; the usual environment-induced decoherence paradigm provides an objective understanding of entropy and of entropy-increase but only for open systems; our matter-gravity entanglement hypothesis provides an extension and completion of the environment paradigm which seems to be capable of offering an objective definition of entropy and an objective explanation of entropy increase also for closed systems.

Where our proposal most sharply distinguishes itself from the traditional account is in its description of the equilibrium states of a black hole in a (say, spherical) box.  Traditionally \cite{HawkingEuc} such a state is modelled as a -- highly impure --  total thermal state of matter and gravity at the Hawking temperature.   On our proposal it is a total pure state of matter and gravity which is entangled in just such a way that the partial state of the matter alone as well as the partial state of the gravity alone are each approximately thermal (again, at the Hawking temperature).   In the traditional understanding of the equilibrium states of quantum black holes, it has never been clear what is the relationship between the entropy of the black hole itself (which one supposes is made out of pure gravitational field) and the entropy of the black hole's thermal atmosphere (which is mostly\footnote{We say `mostly' here because a small part of the thermal atmosphere will, of course, consist of gravitons -- and we assume that these, in turn, also only contribute in a small way to the full description of the state of the gravitational degrees of freedom when a black hole is present.} matter).   In particular it was unclear whether the total entropy should be equated with the former, or the latter, or the sum of the two.   And while there were a number of indications \cite{Page, Wald} that the black hole entropy and the thermal atmosphere entropy had the same value\footnote{Prior to the string theory work discussed below, the former was strongly suggested by the fact (see the discussion in Endnotes (i) and (iii) in \cite{KayAbyaneheeee}) that the entropy is equal to the entropy, $S$, derived from the standard equilibrium statistical mechanical formula  $S=k(\log Z-{\beta\partial/\partial\beta}\log Z)$ from the Gibbons-Hawking Euclidean quantum gravity partition function \cite{GibbonsHawking}, (in Planck units) $Z(\beta) = e^{-\beta^2/16\pi}$, for pure (i.e.\ matterless) gravity in a spherical box;  the latter is strongly suggested by the success of the `t Hooft brick wall model \cite{tHooft, MukohyamaIsrael} which indeed accounts for black hole entropy entirely in terms of the thermodynamic entropy of the thermal atmosphere}, there was no understanding of why this should be the case. We call this the \textit{thermal atmosphere puzzle}.  On our proposal, it would clearly be wrong to add them and natural that they should be equal in the sense that the von Neumann entropy of the reduced state of the gravitational field necessarily equals the von Neumann entropy of the reduced state of the (matter part of the) thermal atmosphere and both are necessarily equal to the matter-gravity entanglement entropy.  We remark that, as we anticipated above, this satisfactory resolution of the  thermal atmosphere puzzle relies on the fact that our hypothesis (i.e.\ of matter-gravity entanglement) is based on the factorization of the total Hilbert space as a tensor product between matter and gravity rather than some other factorization such as into the tensor product of a Hilbert space for matter except for the electromagnetic field and a Hilbert space for gravity together with the electromagnetic field.  

Recently we have explored how our very different understanding of black hole equilibrium states can be reconciled with work of string theory related to black hole entropy which, in its present versions, seems to presuppose the traditional view.    Our results so far \cite{KayModern, KayStringy, KayThermality}  show some promising indications that not only can they indeed be reconciled, but that a clearer understanding of black hole entropy and a clearer resolution of the information loss puzzle emerges once the string theory results are reworked and reinterpreted so as to be compatible with our proposal.   Further results \cite{KayOrtiz, KayEnclosed, KayLupo} suggest a different from usual interpretation of the AdS/CFT correspondence according to which it is a bijection between the boundary CFT and a subtheory of the bulk theory consisting of just its matter degrees of freedom.  To end this essay, we outline the main ideas and results of each of these pieces of work.

{\bf Our first set of results} concerns the impressive quantitative agreement between the results of Strominger and Vafa \cite{StromingerVafa} and subsequent authors for the entropy of extremal and near-extremal black holes and the original Hawking entropy formulae and also between the semi-qualitative results of Susskind \cite{Susskind} and of Horowitz and Polchinski \cite{HorowitzPolchinski, HorowitzChandra} for the entropy of, say, Schwarzschild black holes and the original Hawking entropy formula for those.   These results clearly indicate that string theory is capable of providing an understanding of black hole entropy.  But there are unsatisfactory puzzling issues too:   Strominger and Vafa obtain the entropy as the logarithm of the degeneracy of an energy-level.  Yet (to quote our paper \cite{KayModern}) the degeneracy of the $n$th energy level of the textbook Hydrogen atom Hamiltonian is $n^2$ but we would not conclude that the Hydrogen atom has an entropy of $k\log n^2$!  There is a related unsatisfactory puzzling issue in the work of Susskind and of Horowitz and Polchinski.   They derive the entropy of a Schwarzschild black hole up to a small unknown constant with an argument which we now sketch.   (In what follows, we take $\hbar$ and $c$ to equal 1.   Following \cite{HorowitzPolchinski, HorowitzChandra} we assume we can work with (1+3)-dimensional strings;  $\ell$ stands for the string length scale, $g$ for the string coupling constant and $G$ for Newton's constant, related to $g$ and $\ell$ by $G=g^2\ell^2$.)    Horowitz and Polchinski assume that, as one scales $\ell$ up and $g$ down from their physical values, keeping $G=g^2\ell^2$ fixed, a Schwarzschild black hole of mass $M$ will go over to a long string with roughly the same energy, $\epsilon= M$. The density of states of such a long string, for small enough $g$, is known, very roughly (i.e.\ omitting an inverse-power prefactor) to take the exponential form, $\sigma_{\mathrm{ls}}(\epsilon) = C_{\mathrm{ls}}e^{\ell\epsilon}$
($C_{\mathrm{ls}}$ a constant with the dimensions of inverse energy of the same order of magnitude as $\ell$).  Horowitz and Polchinski then say that the `logarithm' of this is approximately given by $\ell\epsilon$ and propose that $k$ times this should
be equated with the entropy, $S$ of a (Schwarzschild) black hole provided that one does the equating when, to within an order of magnitude or so, $\ell = GM$. Combining these latter two equations (and replacing $\epsilon$ by $M$) they arrive at the
conclusion that the entropy of the black hole will be a moderately sized constant times $kGM^2$ which agrees, up to an undetermined value for the constant, with the Hawking value, $4\pi kGM^2$ for the entropy of a black hole.

The unsatisfactory puzzling issue in this apparent derivation of black hole entropy is that it is, of course, not really meaningful to take the logarithm of such a (dimensionful!) quantity.  Really, before one takes the logarithm, one would need to multiply $\sigma_{\mathrm{ls}}(\epsilon)$ by a constant with the dimensions of energy but, in \cite{HorowitzPolchinski, HorowitzChandra} no such constant is provided by the theory.  

What we propose in \cite{KayModern}  (see also \cite{KayThermality}, \cite{KayStringy}) is that the Horowitz-Polchinski scenario be replaced by a scenario in which, as one scales the string length scale, $\ell$, up and the string coupling constant, $g$, down from their physical values, keeping $G = g^2\ell^2$ fixed, an \textit{equilibrium state} consisting of a (4-dimensional) Schwarzschild black hole of mass $M$ in contact with its (mostly
matter) atmosphere in a box of given total energy, $E$, will go over to an equilibrium state of similar total energy, $E$, consisting of a single long string, with mean energy, $\bar\epsilon$ of a similar magnitude to $M$, in contact with an atmosphere of small strings in a, suitably rescaled, box. If we ignore certain inverse-power prefactors, each of the long string and the stringy atmosphere densities of states, which we call $\sigma_{\mathrm{ls}}$ and $\sigma_{\mathrm{sa}}$, will take the form $Ce^{\ell\epsilon}$ (with different values, $C_{\mathrm{ls}}$ and $C_{\mathrm{sa}}$, say).  

We then appeal to a separate piece of work \cite{KayThermality} on the foundations of statistical mechanics (which we carried out partly in preparation for the analysis of this string theory scenario and which we will briefly outline below) to conclude that, if we regard the total system consisting of the long string weakly coupled to its stringy atmosphere to be in a pure total quantum state which is chosen at random from the set of all possible pure states with energy in a narrow band around $E$, then their reduced states will highly probably be very close (in a certain sense which is explained in \cite{KayThermality}) to approximately thermal states at inverse temperature $k\ell$, each with mean energy very close to $E/2$, while the entanglement entropy between the long string and the atmosphere of small strings will highly probably be very close to $k\ell E/4$ (up to a small logarithmic correction).   We then replace the Horowitz-Polchinski assumption by the assumption that when one scales things back, the total pure state of the long string/string atmosphere system goes over to a pure state of the black hole/atmosphere system and we can equate this mean energy with a constant of order 1 times the black hole mass $M$ and also equate this entanglement entropy with the entanglement entropy between the black hole and its atmosphere provided we do the equating when, to within an order of magnitude or so, $\ell = GM$ -- and also that the approximately thermal reduced states of long string and stringy atmosphere go over to approximately thermal reduced states of the black hole and its atmosphere at the same temperature.   In this way, we arrive at the conclusion that our black hole equilibrium (total pure!) state has the property that the reduced states of the black hole (i.e.\ most of the gravity) and of its (mostly matter) atmosphere are each thermal at inverse temperature a constant of order one times $kGM$ while their entanglement entropy (by our matter-gravity entanglement hypothesis, approximately the physical entropy) is a constant of order one times $kGM^2$.  In fact (but see \cite{KayModern, KayStringy} for further discussion of the significance of this) by equating $\ell$ with $8\pi GM$ one obtains exactly Hawking's formula, $1/T = 8k\pi GM$ for the inverse temperature and exactly Hawking's formula, $S=4k\pi GM^2$, for the entropy.  

Our paper \cite{KayThermality} considers a very general setting in which one has a total closed system (we use the term `totem' for short) consisting of two weakly coupled quantum systems (called `system' and `energy bath') with prescribed densities of states, $\sigma_{\mathrm{sys}}$ and $\sigma_{\mathrm{bath}}$, and investigates what can be said about the reduced state of the system when the totem is in a random pure state with energy in a narrow band around some fixed energy $E$.   In the earlier paper \cite{GoldsteinEtal}, Goldstein et al.\  had considered such a setting in the case that the system is much smaller than the energy bath and showed that, with high probability, the system will find itself in a thermal state (at inverse temperature 
$k d\sigma_{\mathrm{bath}}(\epsilon)/d\epsilon$).   This provided an attractive replacement for the traditional explanation of the thermality of a small system in contact with an energy bath based on the assumption that the totem is in a microcanonical ensemble.  By considering, instead, the totem state to be a pure state (randomly chosen with energy in a narrow band) one obtains foundations for statistical mechanics which are more on a par with the usual foundational assumption of quantum mechanics that the state of a closed system be pure. \cite{KayThermality} goes beyond the work in \cite{GoldsteinEtal} by considering the case where system and energy bath are of comparable size.  It shows that (as it also shows to be the case for a total microcanonical ensemble) for general densities of states (and in particular, for densities of states which grow as a large power of the energy as one expects to hold approximately for ordinary non-gravitational physical systems) the reduced state of the system (or of the energy bath) will not necessarily be thermal but, whether or not it is thermal, the values of the mean energy as well as of the entropy and of other thermodynamic quantities of our system will, with high probability, hardly depend on the particular random pure state one chooses for the totem.    Moreover, it provides a formula for a universal density operator for the system\footnote{$\rho_{\mathrm{S}}^{\mathrm{modapprox}}$  is universal in the sense that the formula for it involves certain quantities (called $|\widetilde{\epsilon, i}\rangle$) which depend on the exact choice of the total pure state, but are such that the value of the system's mean energy or entropy etc.\ don't depend on the values of these quantities}, $\rho_{\mathrm{S}}^{\mathrm{modapprox}}$ (also reproduced in \cite{KayModern}) which, if used to compute quantities such as the system's mean energy or entropy gives values close to those one obtains for the vast majority of our random totem pure states.  (And there is, of course, a similar universal density operator, $\rho_{\mathrm{B}}^{\mathrm{modapprox}}$,  for the energy bath.) Moreoever, in the special case that the densities of states rise exponentially with energy, $\rho_{\mathrm{S}}^{\mathrm{modapprox}}$, and its counterpart for the energy bath, turn out to be approximately thermal in a certain sense.  We remark that the value one obtains with $\rho_{\mathrm{S}}^{\mathrm{modapprox}}$ for the system entropy is (of course) the same as the value one obtains with 
$\rho_{\mathrm{B}}^{\mathrm{modapprox}}$ for the energy bath entropy and is of course the same as the system-energy bath entanglement entropy.

The formalism of \cite{KayThermality}, and in particular the appropriate $\rho_{\mathrm{S}}^{\mathrm{modapprox}}$  is what we used in \cite{KayModern} and \cite{KayStringy} in order to compute the mean energy and the entanglement entropy quoted above -- by identifying the system with the long string and the energy bath with the atmosphere of small strings.

In conclusion, we obtain the same semi-qualitative results as Horowitz and Polchinski with their entropy for a lone black hole replaced by our matter-gravity entanglement entropy of our black hole equilibrium state in a box.   Our scenario, though, is free from the unsatisfactory puzzling issue we mentioned above; in a sense, our proposal supplies the missing constant with the dimensions of energy.   In the paper \cite{KayStringy}  we do a more sophisticated analysis with a more realistic string-theory density of states involving a suitable inverse power prefactor.  This both explains why (for suitable ranges of the relevant parameters) there are multistring equilibrium states which consist of one single long string and an atmosphere of small strings and leads to the same qualitative conclusions regarding the values for the matter-gravity entanglement entropy and for the reduced temperatures.

{\bf Our second set of results} concerns the AdS/CFT correspondence \cite{Maldacena1} which is usually thought to be a full equivalence between a quantum gravity theory in the bulk of Anti de Sitter space (AdS) and a conformal field theory (CFT) on the AdS conformal boundary.   By considering states of quantum gravity which contain black holes and which are modelled classically by the Schwarzschild-Anti de Sitter (Schwarzschild-AdS) spacetime, and by arguing that it is correct to describe these states as in our above discussed description of black hole equilibrium states in terms of a pure total state, we have argued in \cite{KayOrtiz} and \cite{KayEnclosed} that the AdS/CFT correspondence is, instead, a bijection between the boundary CFT and just the matter degrees of freedom of the bulk AdS quantum gravity theory.   As explained in those papers, this seems to offer a resolution to a puzzle raised \cite{ArnsdorfSmolin} by Arnsdorf and Smolin:  The puzzle arises, if one adopts the usual view of full equivalence, from the fact that Rehren has shown in \cite{Rehren1, Rehren2} that any CFT on the conformal boundary of AdS is also equivalent, under a natural form of fixed-background holography which he introduced in these papers, to a quantum field theory on the AdS bulk (satisfying vanishing boundary conditions at the conformal boundary and) obeying an appropriate version of commutativity at spacelike separation.   Such a commutativity condition would seem to be appropriate for a bulk theory involving matter, but not for one involving gravity.   

A key part of our discussion in \cite{KayOrtiz} centers around the question:\  What becomes of a (non-gravitational) quantum field theory on a fixed Schwarzschild-AdS spacetime background when one switches on gravity?   As is well-known,  the maximally extended such classical Schwarzschild-AdS spacetime has a quadruple-wedge structure similar to that of the Kruskal spacetime. If one studies quantum field theory in this fixed background, it is straightforward and standard that there will be a pure  Hartle-Hawking-Israel-like state \cite{HartleHawking, Israel, Sanders} which is entangled between the left and right wedge in just such a way as to be thermal on each wedge separately.  In \cite{Maldacena2}, Maldacena assumes that the full state of quantum gravity, say in the right wedge, will be similarly thermal and similarly entangled with a similar thermal state in the left wedge.   But in \cite{KayEnclosed} (see also \cite{KayLupo}) I argue (see below for an outline of the argument) that, once one switches on the dynamical gravitational field, the horizon becomes unstable and the right wedge becomes a full quantum spacetime\footnote{We refer to the right wedge here as a `quantum spacetime' because we expect the classical describability to break down near where the horizon used to be.} in its own right with an overall pure state of matter and gravity which -- in line with the understanding of black hole equilibrium states that we propose here -- is entangled between matter and gravity in just such a way that each of matter and gravity separately are approximately thermal.   Interestingly, recently, a number of other authors  \cite{AveryChowdhury, Mathur, Chowdhury, ChowdhuryParikh} have argued on quite different grounds internal to string theory that the right wedge becomes a full quantum spacetime in its own right\footnote{Note that one can also similarly argue that the left wedge becomes a full quantum spacetime in its own right. The authors of \cite{AveryChowdhury, Mathur, Chowdhury, ChowdhuryParikh} differ from us in that they still consider overall pure quantum states on the union of right and left wedges which are entangled between these two now disconnected quantum spacetimes, whereas our conclusion is that, in physically relevant states, such a left wedge would neither be geometrically connected to, nor quantum mechanically entangled with, the right wedge and may as well be considered not to exist.}

It is not in doubt that, under the AdS/CFT correspondence, the CFT on the conformal boundary of the right wedge is in a thermal state with a (von Neumann) entropy equal to the entropy of the quantum gravity theory on the right wedge.   But the standard intepretation of this (including the interpretation implicit in \cite{Maldacena2})  is that the quantum gravity theory on the right wedge is also in a (total) thermal state (due to entanglement with the left wedge) and its entropy is the von Neumann entropy of this total thermal state.  On our view, the right wedge has become a quantum spacetime in its own right and the state of quantum gravity on it is a pure state, but what one means by its entropy is not the von Neumann entropy of this total pure state (which is of course zero) but its matter-gravity entanglement entropy --  i.e.\ the von Neumann entropy of just the matter (also of just the gravity but that's by the by).   Our argument for the AdS/CFT correspondence being a bijection between the boundary CFT and just the matter degrees of freedom of the bulk theory is that this would naturally fit with the equality of the latter von Neumann entropy of the bulk matter with the von Neumann entropy of the thermal state of the boundary CFT.  As a reasonableness check on this, we verified, in \cite{KayOrtiz} in a simple linear scalar field model (with vanishing boundary conditions on the conformal boundary) on 1+1 and 1+2 dimensional analogues to the geometry of the Schwarzschild-AdS right wedge that (when both are suitably regularized by regularizing the bulk entropy with the brick-wall model of \cite{tHooft, MukohyamaIsrael}) the entropy of the boundary CFT according to fixed background holography is the same as the entropy of the bulk scalar field when the latter is in the Hartle-Hawking-Israel state.

\begin{figure} 
   \centering
    \includegraphics*[scale=0.45, trim=100 100 0 0, clip]{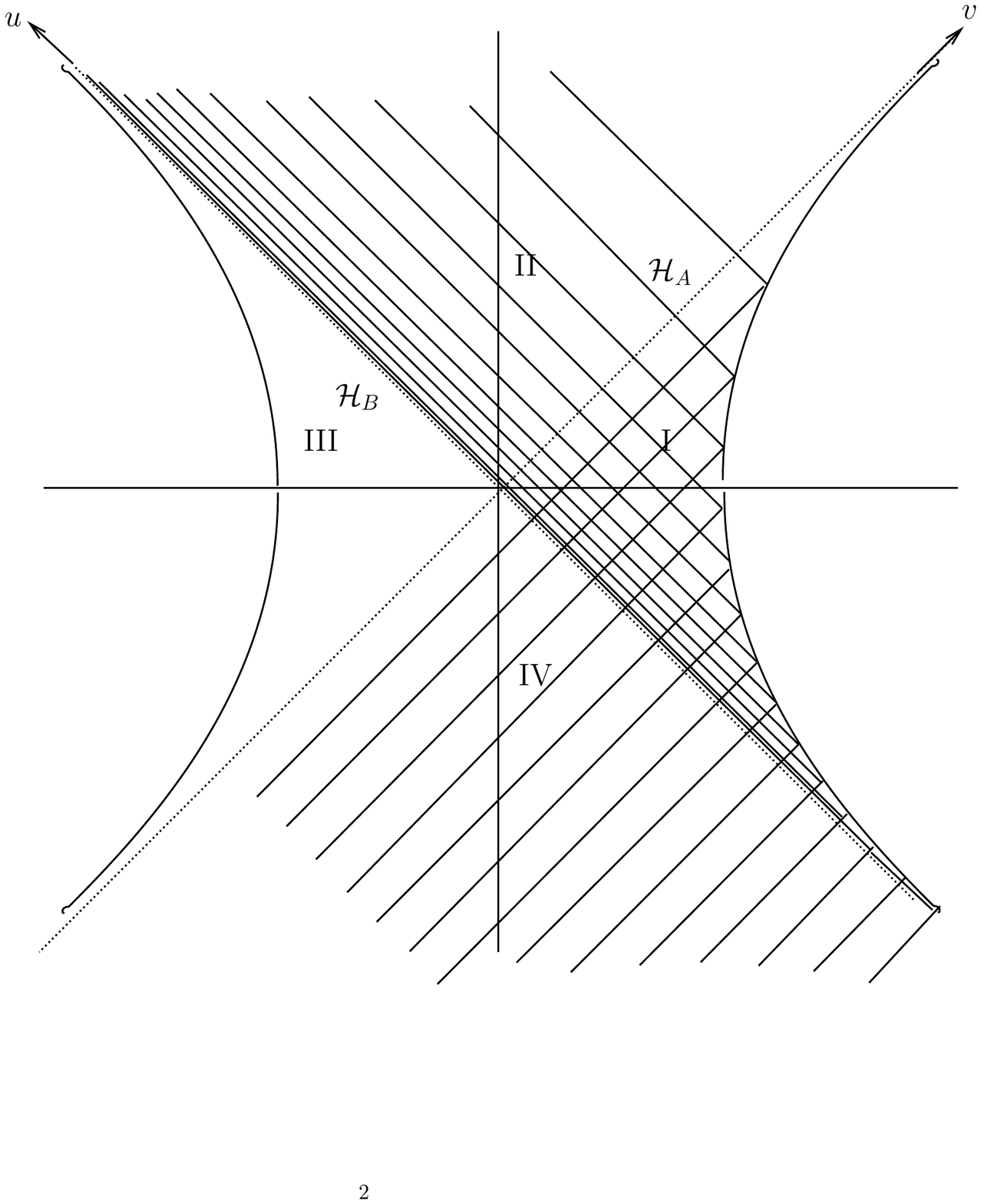}
   \caption{(= Figure 3 of Reference \cite{KayThermality}.  Reproduced here with kind permission from Springer Science + Business Media.) Schematic diagram of the four wedges of the region of 1+1 Minkowski space between the two components of a hyperbolic boundary (i.e.\ the curve $uv= -1$, in the indicated double-null coordinates, $u$ and $v$) which may be thought of as a pair of accelerated mirrors.  Shown are lines of constant phase of (the restriction to Region IV of) an initially right-moving plane wave.  The wave reflects off the mirror in Region I and so do its lines of constant phase, piling up towards the horizon, ${\cal H}_B$ ($v=0$).  We argue that a similar pile-up occurs in the Schwarzschild-AdS spacetime leading to the instability of the ${\cal H}_A$ and ${\cal H}_B$ horizons there.
\label{fig3}}
 \end{figure}

It remains for us to explain why we believe the horizons of Schwarzschild-AdS to be unstable.  A strong clue towards this is already given by the fact -- evident from the quadruple-wedge geometry of the Schwarzschild-AdS spacetime --  that (assuming that past-directed light rays which hit the conformal boundary reflect off it in the obvious way) if an observer were to pass from (say) the past wedge to (say) the right wedge, then just after they cross the past horizon of the right wedge, they will see an infinite amount of the history of the past wedge in a finite amount of time.  This is strongly reminiscent of the celebrated fact \cite{HawkingEllis, SimpsonPenrose} that if an observer crosses the Cauchy horizon of the (non-extremal) Reissner-Nordstr\"om spacetime then they will see an infinite amount of history in a finite amount of time just before they cross it.  The latter observation is well-accepted as indicating the instability of that Cauchy horizon (for recent results on this, see \cite{Dafermos2003, Dafermos2004}) and similarly (although, for a number of reasons which are explained in \cite{KayEnclosed}, the analogy is not an exact analogy) one expects the above fact about Schwarzschild-AdS to indicate instability of its horizons.   Our main argument that there is actually such an instability concerns a simple analogue system:\ the massless Klein Gordon equation on the region of 1+1 Minkowski space in between the two branches of the hyperbola (in the usual Minkowski coordinates $x$ and $t$ and taking the speed of light, $c$, to equal 1) $x^2-t^2=1$ -- in double-null coordinates ($u=t-x$, $v=t+x$) the curve $uv=-1$ -- see Figure \ref{fig3}.   We can think of the two branches of the hyperbola as a pair of accelerating (/decelerating) mirrors and we assume the physical effect of these on our field is to impose vanishing bondary conditions.   These mirrors are analogous in an obvious way to the disconnected conformal boundary of the Schwarzschild-AdS spacetime.   As one sees from Figure \ref{fig3}, an initially right-moving classical plane wave, emerging from the past wedge will reflect off the mirror and its wave fronts will pile up on the horizon ${\cal H}_B$ -- with the result (see \cite{KayEnclosed} for details) that the $vv$-component of its stress-energy tensor becomes infinite there.  In \cite{KayEnclosed} we show that there are also finite energy wave packets with a similar pile-up property and a similar singularity in their stress-energy tensor.  We further show that suitable compactly supported arbitrarily small initial data on a suitable initial surface will develop an arbitrarily large stress-energy scalar near where the two horizons cross.  For the quantum theory, we show that while there is a regular Hartle-Hawking-Israel-like state, there are coherent states built on this (whose expectation values are the above sorts of classical solutions) for which there is a similar singularity in the expectation value of the renormalized stress-energy tensor.   We conjecture that similar results hold for the Schwarzschild-AdS spacetime (in any dimension) and that they entail the sort of horizon instability we referred to above.

\end{document}